\documentclass[manuscript=article]{achemso}

\usepackage{amssymb}
\usepackage{amsmath}
\usepackage{color}
\usepackage{subfigure}
\usepackage{graphicx}

\usepackage[version=3]{mhchem} 
\usepackage{natbib}
\usepackage{mciteplus}

\setkeys{acs}{usetitle = true}

\def\@dotsep{4.5}
\makeatother

\usepackage{bm}
\usepackage{graphicx}
\usepackage{amssymb}
\usepackage{amsmath}
\usepackage{hyperref}

\title{Three-Dimensional Wave Packet Approach for the Quantum Transport
of Atoms through Nanoporous Membranes }

\author{Alfonso Gij{\'o}n}
\affiliation{Instituto de F\'{\i}sica Fundamental,
Consejo Superior de Investigaciones Cient\'{\i}ficas (IFF-CSIC), Serrano 123,
28006 Madrid, Spain}

\author{Jos{\'e} Campos-Mart\'{\i}nez}
\affiliation{Instituto de F\'{\i}sica Fundamental,
Consejo Superior de Investigaciones Cient\'{\i}ficas (IFF-CSIC), Serrano 123,
28006 Madrid, Spain}

\author{Marta I. Hern\'{a}ndez} \email {marta@iff.csic.es}
\affiliation{Instituto de F\'{\i}sica Fundamental,
Consejo Superior de Investigaciones Cient\'{\i}ficas (IFF-CSIC), Serrano 123,
28006 Madrid, Spain}

\date{\today}

\keywords{ graphynes, graphdiyne, two-dimensional materials, nanofiltration,
ab initio calculations }

\begin{document}

\maketitle

\begin{abstract}
 
Quantum phenomena are relevant to the transport of light atoms and molecules
through nanoporous two-dimensional (2D) membranes. Indeed, confinement
provided by (sub-)nanometer pores enhances quantum effects such as tunneling
and zero point energy (ZPE), even leading to quantum sieving of different
isotopes of a given element. However, these features are not always taken into
account in approaches where classical theories or approximate quantum models
are preferred. In this work we present an exact three-dimensional wave packet
propagation treatment for simulating the passage of atoms through periodic 2D
membranes. Calculations are reported for the transmission of $^3$He and $^4$He
through graphdiyne as well as through a holey graphene model. For
He-graphdiyne, estimations based on tunneling-corrected transition state
theory are correct: both tunneling and ZPE effects are very important but
competition between each other leads to a moderately small $^4$He/$^3$He
selectivity. Thus, formulations that neglect one or another quantum effect are
inappropriate. For the transport of He isotopes through leaky graphene, the
computed transmission probabilities are highly structured suggesting
widespread selective adsorption resonances and the resulting rate 
coefficients and selectivity ratios are not in agreement with predictions 
from transition state theory. Present approach serves as a benchmark
for studies of the range of validity of more approximate methods.  

\end{abstract}

\maketitle

\vskip 1.cm
KEYWORDS: quantum sieving, nanoporous two-dimensional materials, Helium
isotopes, graphdiyne, zero point energy, tunneling

\newpage

 Recent progress in the fabrication of nanoporous two-dimensional (2D)
 membranes has led to propose them as efficient sieves at the molecular
 level\cite{Nature-nano:2012,Marlies:2013,review-JPCL:2015}. Particularly, it
 has been suggested that these  membranes could be used for the separation of
 a specific isotope within an atomic or molecular
 gas\cite{Schrier:10,Hauser:2012}, a process which is both important and
 challenging. For example, $^3$He is essential to several applications ranging
 from security to basic research but it is very rare and its growing demand is
 leading to an acute shortage of this
 species\cite{ScienceHe3:09,Nature-He:2012}. Moreover, separation of $^3$He
 from the much more abundant $^4$He -in turn, commonly extracted from natural
 gas- usually involves very expensive cryogenic methods. Separation of the
 heavier isotopes of H$_2$ is also crucial for various
 technologies\cite{RSCAdv:2012}. Ideally, single-layer membranes should 
 involve a low energy consumption and efficiency, provided that the
 pores are designed to optimize the desired separation process. 

 For sufficiently low pressures, the study of purification of a gas mixture
 can be modeled by the dynamics of an atom or molecule passing through a
 pore of the 2D membrane. In this way, rate coefficients for the transmission
 of the isotopic species of the mixture, say $k_a$ and $k_b$, are
 independently computed at a given temperature $T$, and the efficiency for 
 the isotope separation is estimated by means of the selectivity ratio
 $S_{a/b}(T) = k_a(T)/k_b(T)$. Since isotopes are chemically identical, the 
 separation mechanism must be provided by mass-dependent
 dynamical effects. At sufficiently low temperatures, quantum effects may
 entail large selectivity ratios as compared with those based on classical
 diffusion, for instance. Recently, a rather large number of works have shown
 that quantum tunneling  might rule an efficient separation
process\cite{Hauser:2012,Schrier:12,ceotto:2014,ourJPCC2014,Lalitha2015,Li:2015,Qu2016}.  
However, these studies are based on one-dimensional (1D) quantum-mechanical
calculations, whereas the molecular motion actually occurs in the
three-dimensional (3D) space where differences between the quantum energy
levels of the isotopes confined within the pores can also lead to quantum
sieving\cite{Beenakker:95}. These effects, which may be generally termed as
zero point energy (ZPE) effects\cite{Note-zpe}, have been invoked in various
theoretical\cite{hankel-pccp:2011,Bathia:2005,Schrier-cpl2012,Marlies:2013} 
and experimental\cite{Zhao:06,Nguyen2010} works on porous materials. 
 It is worth to emphasize that tunneling and ZPE effects work in opposite
 directions: while tunneling favors the passage of lighter molecules, ZPE
 increases the transmission of the heavier ones since their smaller ZPE
 is associated to a smaller ``effective size''. Therefore, it is important to
 evaluate the relative role of each effect on the systems of interest. 

We have recently studied\cite{ourJPCA:2015} the interplay between tunneling
and ZPE effects in the transmission of He isotopes through
graphdiyne\cite{graphd-chemcomm:2010,graphd-jacs:2015}, a promising new
material for molecular separation
applications\cite{graphd-chemcomm:2011,nanoscalemitbis:2012,jpclours:2014,ourJPCC2014}
as it exhibits regularly distributed pores of sub-nanometer size. To this
end we relied on transition state theory (TST) with especific inclusion of
reaction-path tunneling corrections. He-graphdiyne interactions were
represented by a force field validated from {\em ab initio} electronic
structure calculations\cite{ourJPCC2014}. It was found that, if only tunneling
effects are considered (ignoring ZPE), $^3$He transmission rate is larger than
the $^4$He one in the studied 20-100 K temperature range, reaching a
selectivity factor of $\approx$ 2.5 at 20 K. This result is similar to
findings accomplished from 1D calculations on related
systems\cite{Schrier:12}. However, the complete theory, which also includes 
ZPE effects, leads to a {\em qualitatively} different conclusion: 
transport becomes more probable for $^4$He and  at 20 K the selectivity ratio
is also  $\approx$ 2.5 but this time favoring $^4$He instead of $^3$He. It
became clear that the transmission of atoms through the membrane must be
studied within a 3D model and that accurate quantum calculations may be needed
in order to properly account for the delicate competition between the above
mentioned quantum effects.

Here we present an accurate quantum-mechanical formulation to study
the transmission of an atom through a periodic 2D membrane, therefore going  
beyond TST. The time-dependent Schr{\"o}dinger equation is solved by 
propagating 3D wave packets and, from the calculation of the flux through a
surface separating the incident and transmitted wave packets, transmission
probabilities and rate coefficients are obtained. Time-dependent
quantum-mechanical methods have been extensively applied to various
collisional and photodissociation processes\cite{reviewtdwp:97,BKurti:08} but,
although one of their first applications was devoted to the scattering of atoms by 
surfaces\cite{Yinnon:83}, we are not aware of a generalization of this approach to
the scattering by a porous membrane. Present 3D  wave packet  method (WP3D in
short) is applied to He-graphdiyne\cite{ourJPCC2014} as well as to a holey
graphene model\cite{CSun:2014}. The goal of this work is to provide
with a trustworthy method for the investigation of quantum phenomena in the
transport of atoms through membranes, in this way allowing us the assessment
of more approximate treatments. It is found that TST is reliable for
He-graphdiyne but not for He-holey graphene. Furthermore, it turns up that in
addition to tunneling and ZPE effects, selective adsorption
resonances\cite{LJD:1936,SMiret:07,NuestroPRB:94} (another genuine quantum
feature) can also play a role in these processes. It is expected that this
research will serve to uncover new clues for the design of optimal pores for
quantum sieving.      

The rest of the paper arranges as follows. First we give the theory for the
transmission of a 3D wave packet through a periodic membrane, accompanied by a
refresher outline of TST. Results are presented and discussed first for
He-graphdiyne, followed by He-holey graphene. The report ends with a
conclusion paragraph.

\vspace{0.5cm}

{\em Theory.}
In the present WP3D approach we consider the scattering of an atom of mass
$\mu$ by a periodic membrane by means of time-dependent 3D wave packet
methods\cite{SplitOperator,K&K:1983}. The membrane coincides with the $xy$
plane of the reference frame, whose origin is set at the center of one of its
pores; hence the position of the atom is given by ${\bf r} = ({\bf R}, z)$,
$z$ being the distance to the membrane plane and ${\bf R} = (x,y)$. The wave
packet representing the atom is discretized on a grid of evenly spaced ${\bf
  r}$ points and at the start of the propagation is given as a product of a
Gaussian wave packet\cite{Heller:75} in $z$ times a plane wave with wave
vector ${\bf K}$ in ${\bf R}$. Thus, as in the
original work by Yinnon and Kosloff\cite{Yinnon:83}, the periodicity of the
system is fully exploited by matching the size of the ($x,y)$ grid to that of
the  unit cell, ($\Delta_x,\Delta_y$), while the values of the 
parallel wave vector ${\bf K}$ are restricted such that the initial plane
wave is commensurate with the membrane lattice. The wave packet is propagated in
time subject to the time-dependent Sch{\"o}dinger equation, using the Split
Operator method\cite{SplitOperator}, and is being absorbed in the
asymptotic regions by  means of a wave packet splitting algorithm\cite{Pernot:91}. 
To obtain the probability of transmission of the atom through the membrane, its
is convenient to write first the asymptotic behavior of the stationary wave
function for a translational energy $E = \frac{\hbar^2 k^2}{2 \mu}$,

\vspace*{-0.75cm}

\begin{eqnarray}
\Psi^+_E ({\bf r}) & 
\left. \begin{array}{c}  \\ \longrightarrow \\ { z \rightarrow \infty}
  \end{array} \right.
 & \sqrt{\frac{\mu}{2\pi \Delta_x\Delta_y
    \hbar^2}} \left[ \frac{e^{i {\bf k} \cdot {\bf r}}}{\sqrt{-k_z}} +
  \sum_{\bf G} A_{\bf G}^{+}  \frac{e^{i \left[k^+_{z,{\bf G}} z + ({\bf K +
          G})  \cdot {\bf R} \right] }}{\sqrt{k_{z, {\bf G}}^+}} \right]
\nonumber \\
    & \left. \begin{array}{c}  \\ \longrightarrow \\ {z \rightarrow -\infty}
  \end{array} \right.
 & \sqrt{\frac{\mu}{2\pi \Delta_x\Delta_y
    \hbar^2}}   \sum_{\bf G} A_{\bf G}^{-}  \frac{e^{i \left[k^-_{z,{\bf G}} z + ({\bf K +
          G})  \cdot {\bf R} \right] }}{\sqrt{-k_{z, {\bf G}}^-}},  
\end{eqnarray}

\noindent
which represents an incident plane wave with a wave vector ${\bf k}= (k_z,\bf
K)$ and a set of reflected ($+$) and transmitted ($-$) waves with amplitudes
$A_{\bf G}^{\pm}$ labeled by the reciprocal lattice vector, ${\bf G}$. Note
that the parallel wave vectors of these waves obey the Bragg condition whereas
the perpendicular one is modified to satisfy the conservation of energy,
$k_{z,{\bf G}}^{\pm} = \pm \left[ k^2  - ({\bf K} + {\bf G})^2 \right]^{1/2}$,
as energy exchange with the membrane is neglected in the present approach.
This function is normalized as $< \Psi^+_E \mid \Psi^+_{E'} ({\bf
  r})> = \delta(E-E')$. It can be shown that the (total) transmission
probability, which is the sum of the squared transmission
amplitudes, can be also obtained from the flux of the stationary wave function
through a surface $z= z_{f}$ separating transmitted from incident and
reflected waves\cite{Miller:74},    

\vspace*{-0.65cm} 

\begin{eqnarray}
P_{trans}(E) & = & \sum_{\bf G} \mid A_{\bf G}^{-}\mid^2 \nonumber \\
         & = & \frac{2 \pi \hbar^2}{\mu} \operatorname{Im} \left( \int dx dy \,
\Psi^{+*}_E(x,y,z_{f}) \frac{d  \Psi^{+}_E}{dz}\mid_{z=z_{f}} \right).
\label{trprob}
\end{eqnarray}

\noindent
We have employed this flux formula for computing $P_{trans}$, where
$\Psi^{+*}_E(x,y,z_{f})$ is obtained from the time-energy Fourier transform
of the evolving wave packet\cite{Zhang:91,cplh2h2:01}. 

The transmission rate coefficient is then obtained from the integration of
$P_{trans}(E)$, properly weighted by the Boltzmann factor: 

\begin{equation}
k(T) = \frac{1}{h  Q_{trans}} \int e^{-E/(k_BT)} P_{trans}(E) dE,
\label{eq4}
\end{equation}

\noindent 
where $Q_{trans}  =  \left( 2 \pi \mu k_B T / h^2 \right)^{3/2}$ 
is the translational partition function per unit volume. In detail,
$P_{trans}$ not only depends on the translational energy but it is also 
labeled by the parallel wave vector ${\bf K}$. A complete 
calculation of the rate coefficients should involve  averaging over a
sufficiently large set of ${\bf K}$ values. In this work we have used initial
wave packets perpendicularly approaching the membrane ({\bf K}={\bf
  0}) and postpone the investigation of effects due the different 
orientations of the incident wave. 

Computational details of the WP3D simulations are provided in the Supporting
Information Section.

Rate coefficients computed in this way are compared with those obtained from
TST as detailed in  Ref.\cite{ourJPCA:2015}. In short, it is assumed that the
reaction path is a straight line perpendicular to the membrane and crossing
the center of the pore, which is the TS. Hence the transmission rate
coefficient can be written
as\cite{hankel-pccp:2011,Truhlar-jcp:72,Truhlar-jpc:79}

\begin{equation}
k_{TST}(T)  =  \gamma \frac{k_B T}{h} \frac{Q^{\ddagger}}{Q_{trans}} f_{tunn}(T),
\label{ktst}
\end{equation}

\noindent
where 

\begin{equation}
Q^{\ddagger} =  \sum_n e^{-E_n/k_BT}
\label{qtst}
\end{equation}

\noindent
is the TS partition function, $E_n$ being the energy levels of the bound states
for the degrees of freedom perpendicular to the reaction path (He in-plane
vibrations inside the pore). In addition, $f_{tunn}(T)$ is a correction for
tunneling effects along the reaction
path\cite{ourJPCA:2015,Truhlar-jcp:72,Truhlar-jpc:79}. Finally, $\gamma$ is a
correction factor (not considered previously) related to the fact that only a
fraction of the membrane is effective for permeation\cite{Wang:2016} and is
defined as  

\begin{equation}
\gamma = n_p \frac{A_{eff}}{A_{uc}}
\label{gamma}
\end{equation}

\noindent
where $A_{eff}$ is the pore effective size, $A_{uc}$ is the area of
the unit cell and $n_p$ is the number of pores per unit cell. Calculation of
$A_{eff}$ is detailed below.

\begin{figure}[h]
\hspace*{-2.cm}\includegraphics[width=8.25cm,angle=0.]{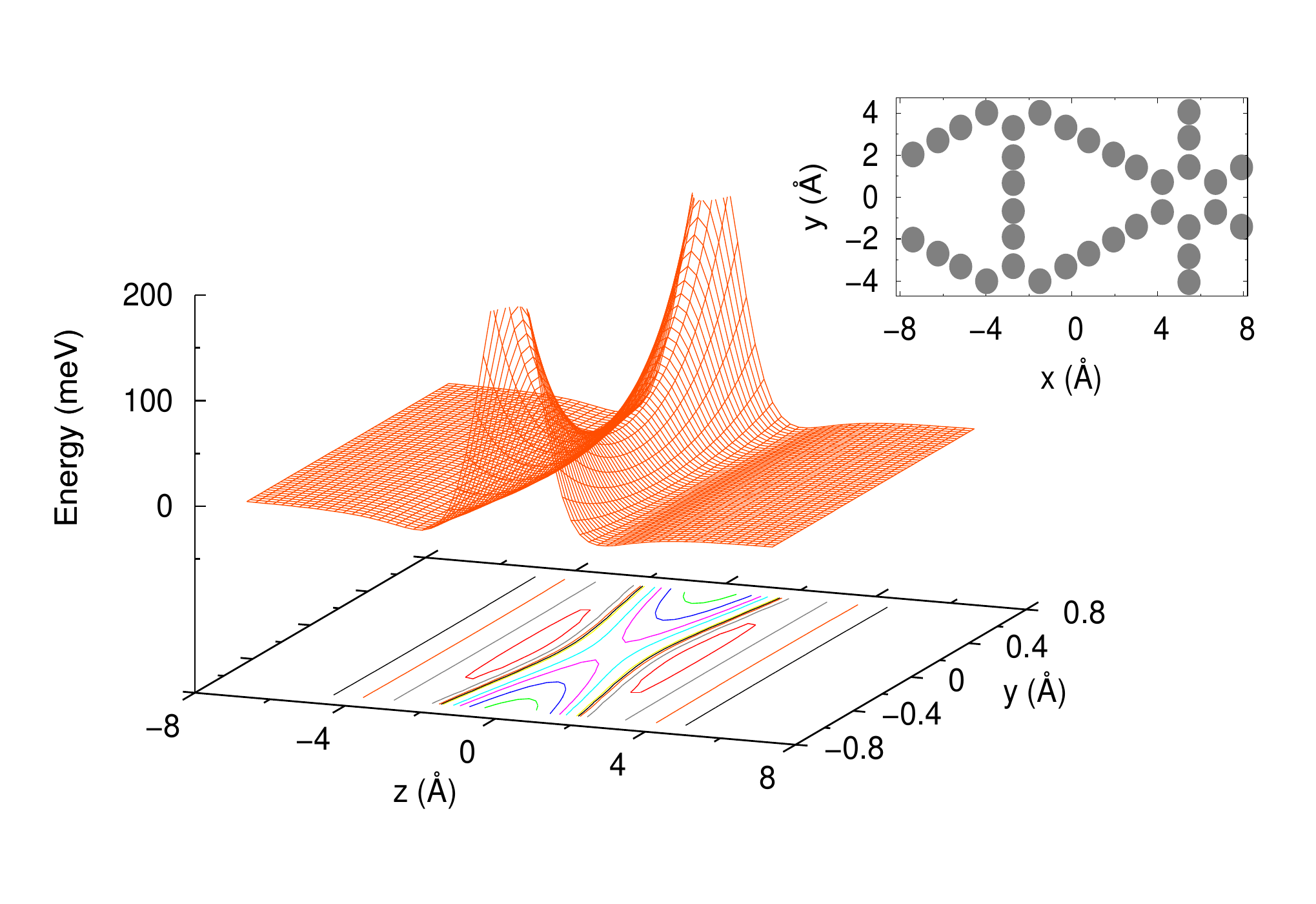}
\caption[] {Right-upper panel: Graphdiyne unit cell employed in the WP3D
  calculations (carbon atoms are depicted by grey filled circles). Left-lower
  panel: He-graphdiyne interaction potential (meV) as a function of the $y$
  and $z$ coordinates, with $x=0$.}
\label{fig1new}
\end{figure}

\begin{figure}[h]
\hspace*{-2.cm}\includegraphics[width=8.25cm,angle=0.]{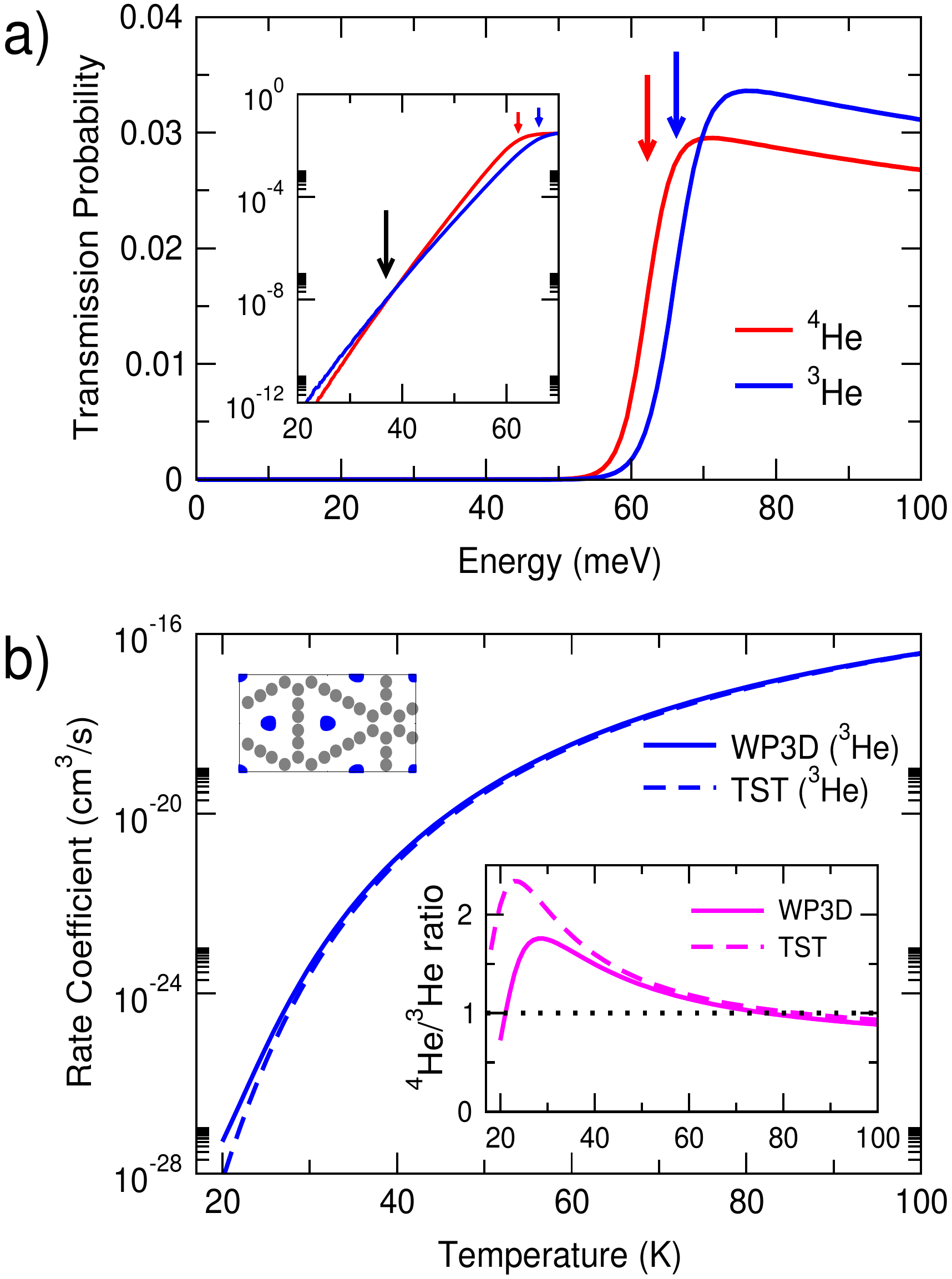}
\caption[] {a) WP3D probabilities of $^4$He and $^3$He transmission 
  through graphdiyne as a function of the translational energy of the atom (in
  meV). Red and blue arrows indicate the $^4$He and $^3$He reaction
  thresholds, respectively, as predicted by TST\cite{ourJPCA:2015}. The inset
  shows these probabilities (in logarithmic scale) for the low energy
  region. The black arrow shows the potential barrier height. b) 
  $^3$He WP3D rate coefficients vs. temperature compared with TST
  estimations. Graphdiyne unit cell, together with the effective area for
  $^3$He transmission as obtained from TST (blue spots), are displayed in the
  upper-left inset. Finally, the $^4$He/$^3$He selectivity vs. temperature is
  presented in the lower-right inset.} 
\label{fig2new}
\end{figure}

\vspace{0.5cm}

{\em Transmission of He isotopes through graphdiyne}. 
Our first choice for performing WP3D calculations is the
transmission of He isotopes through graphdiyne, as for this system a reliable
force field has been already obtained\cite{ourJPCC2014} and TST selectivity
ratios showing an involved behavior have been already
reported\cite{ourJPCA:2015}. Briefly, the He-graphdiyne potential is obtained
as a pairwise sum over He-C pair potentials, the latter 
being represented by an Improved Lennard-Jones (ILJ) formula\cite{ILJ}
whose parameters have been optimized from comparison with benchmark high level
{\em ab initio} calculations\cite{ourJPCC2014}. Graphdiyne unit cell is
depicted in the right-upper panel of Fig.\ref{fig1new} whereas in its
left-lower panel a plot of the He-graphdiyne potential is presented.  The
point ${\bf r}=0$ is a saddle: whereas it corresponds to the maximum of a
barrier potential along the ``reaction-path'' $z$ coordinate -with a height of
$E_0 =$ 36.92 meV- it is a minimum with respect to displacements along the $y$
and $x$ ``in-pore'' degrees of freedom. It is worthwhile to note that 
while the potential barrier is rather low at this saddle point it rapidly
rises for paths different to the minimum energy path. 

Transmission probabilities for $^4$He and $^3$He are presented and compared in
Fig.\ref{fig2new}.a). These probabilities rise around
60 meV, a value much higher than the potential barrier (36.9 meV). The
positions of these thresholds agree quite well with the TST prediction, given
by the lowest energy levels of Eq. \ref{qtst}, 62.2 and 66.2 meV for $^4$He
and $^3$He, respectively\cite{ourJPCA:2015}, and depicted by arrows in the
figure. In addition, it is worth noticing the small value of the probabilities
above threshold: they are slightly lower (higher) than 0.03 for $^4$He
($^3$He). These values can be related to the ratio between the effective size
of the pores and the membrane area, i.e., with the $\gamma$ factor of
Eq.\ref{gamma}. Hence the effective pore size ($A_{eff}$) can be estimated and
it is found  to be slightly lower (higher) than 1.2 \AA$^2$ for
$^4$He ($^3$He). Finally, the behavior below threshold is shown in the inset
of Fig. \ref{fig2new}.a), where it can be seen that the probabilities decrease
exponentially as energy decreases. There the black arrow indicates the
potential barrier. Nearly below this energy $^3$He transmission becomes more
probable, in agreement with previous 1D
calculations\cite{ourJPCC2014,ourJPCA:2015} where, below the barrier, the
lighter atom exhibits a larger tunneling probability.     

Rate coefficients as functions of temperature are determined from these
probabilities (Eq.\ref{eq4}) and the result for $^3$He is presented in
Fig.\ref{fig2new}.b). This WP3D rate coefficient is compared with
that previously reported within TST\cite{ourJPCA:2015} except that in
this work we additionally include the correction given $\gamma$
(Eqs. \ref{ktst} and \ref{gamma}). To that end, we have computed  $A_{eff}$
from the ground state TS wave function, $\Psi_0(x,y)$, as the region where
$\mid\Psi_0(x,y)\mid^2$ is larger than a given cutoff, $f_{cut}$. Taking
$f_{cut}= 10^{-4}$ leads to $\gamma (TST)$= 0.033 and 0.029 for $^3$He and
$^4$He, respectively, a result that nicely matches the values of the
probabilities mentioned above for the two isotopic species. With this choice  
WP3D and TST rate coefficients agree very well along the
whole temperature range, except at the lowest temperatures where the WP3D
coefficients are somewhat larger, probably due to an underestimation of
tunneling from TST. For $^4$He (no shown) WP3D and TST comparison is even more
successful. Finally, the $^4$He/$^3$He selectivity (ratio of rate
coefficients) is reported in the inset of Fig.\ref{fig2new}.b). It can be seen
that there is a fairly good agreement between the WP3D and TST calculations,
although below 40 K WP3D calculations show that the preference for the
transport of the heavier isotope is not as significant as originally predicted
by TST.   

WP3D calculations confirm the conclusions previously drawn from the TST
calculations\cite{ourJPCA:2015}: both tunneling and zero point energy are very
important effects in the permeability of He at low temperatures but, as they
operate in opposite directions, in the end we cannot achieve a large
difference between the transmission rates of the two isotopes.  We would like
to stress that neglect of one or the other quantum effect in the model would
have led to qualitatively erroneous results. Moreover, the very low values of
the related rate coefficients (Fig. \ref{fig2new}.b)) suggest that the actual
flux of these species through the sieve would be extremely
slow. It is well known that permeability usually decreases as selectivity
increases\cite{Robeson:08}. As a possible strategy for achieving 
isotope separation, new  membranes could be designed where one of the two
competing quantum effects are suppressed while the flux of the more permeable
species is kept sufficiently large. With that aim, we report below results for a
model system where tunneling is in principle absent.

\begin{figure}[h]
\hspace*{-2.cm}\includegraphics[width=8.25cm,angle=0.]{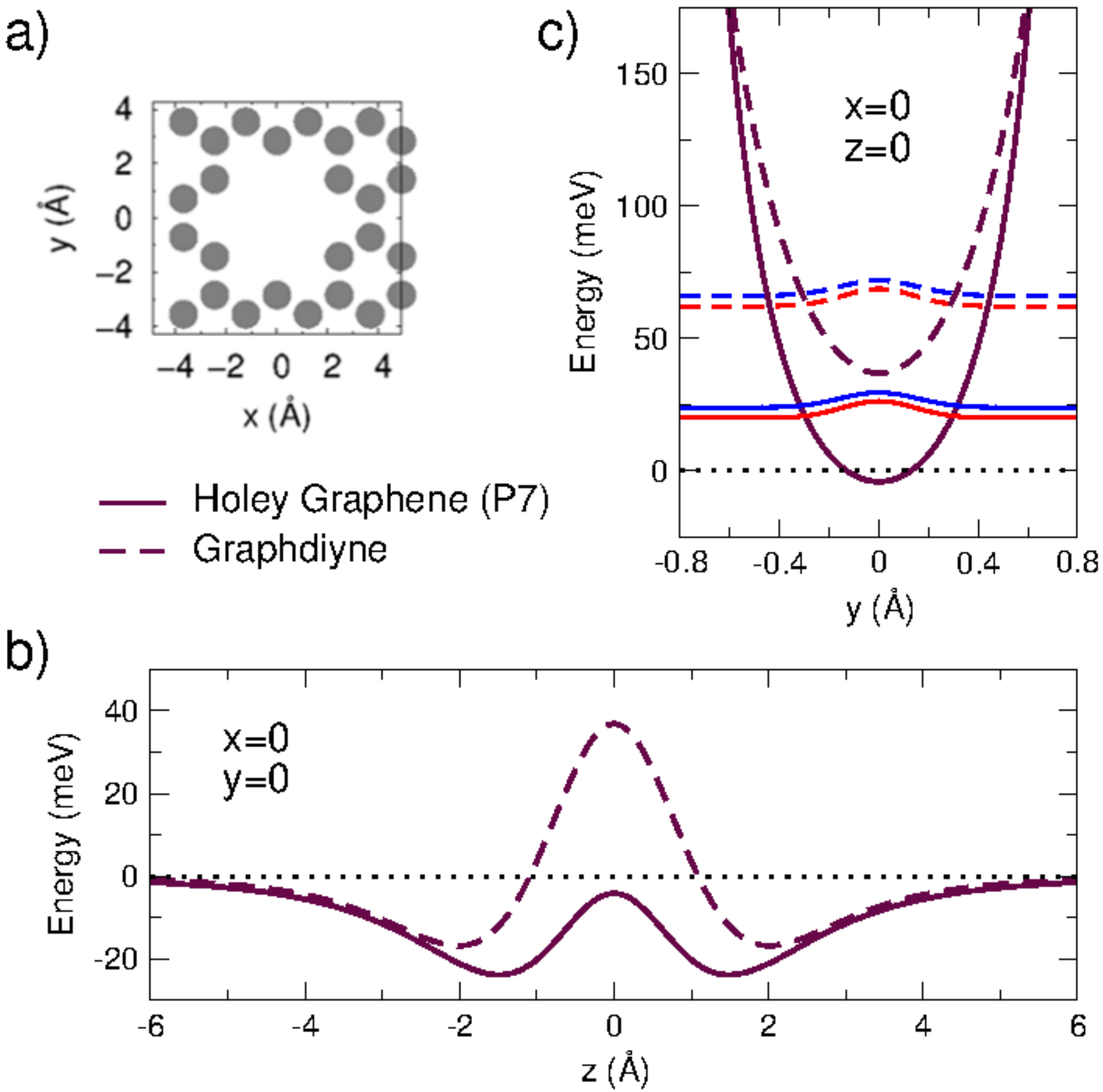}
\caption[] {a) unit cell of the holey graphene model (P7)
  membrane. b) He-P7 interaction potential along the minimum energy
  path for transmission, compared with He-graphdiyne. c) also
  compared with He-graphdiyne, He-P7 ``in-pore'' interaction potential
  (displacement along the $y$ coordinate for $x=0$, $z=0$) and profiles of the
  ground states  at the TS for both He isotopes.}
\label{fig3new}
\end{figure}

\vspace{0.5cm}

{\em Transmission of He isotopes through a holey graphene model (P7)}. 
We have adopted the model of Sun {\em et al}\cite{CSun:2014}, where
nanopores are generated by eliminating atoms from a graphene sheet and a simple
Lennard-Jones pairwise interaction is assumed between He and all the carbon
atoms of the  membrane. Here we have chosen the same pairwise potential
($\sigma=$ 2.971 \AA \, and well depth $\epsilon=$ 1.611 meV) and
a membrane where the pores are created by periodically removing seven rings
from graphene. The unit cell of this membrane, which will be called ``P7
membrane'', is depicted in Fig.\ref{fig3new}.a). Also in that figure the main
features of the He-P7 interaction are compared with those of He-graphdiyne. On
the one hand and in contrast with graphdiyne (Fig.\ref{fig3new}.b)), there is
not a potential barrier along the minimum energy path of He-P7. On the other
hand (Fig.\ref{fig3new}.c)) and now in similarity with the previous membrane,
the difference between the ZPEs of $^4$He and $^3$He is considerable (4
meV). Using TST arguments, one can expect a large $^4$He/$^3$He selectivity
due to this difference as well as the
suppression of tunneling which would favor $^3$He. Interestingly, as
these TS energies are much lower than those of He-graphdiyne, we foresee that
the transmission rates will be much larger in the present system.

\begin{figure}[h]
\hspace*{-2.cm}\includegraphics[width=8.25cm,angle=0.]{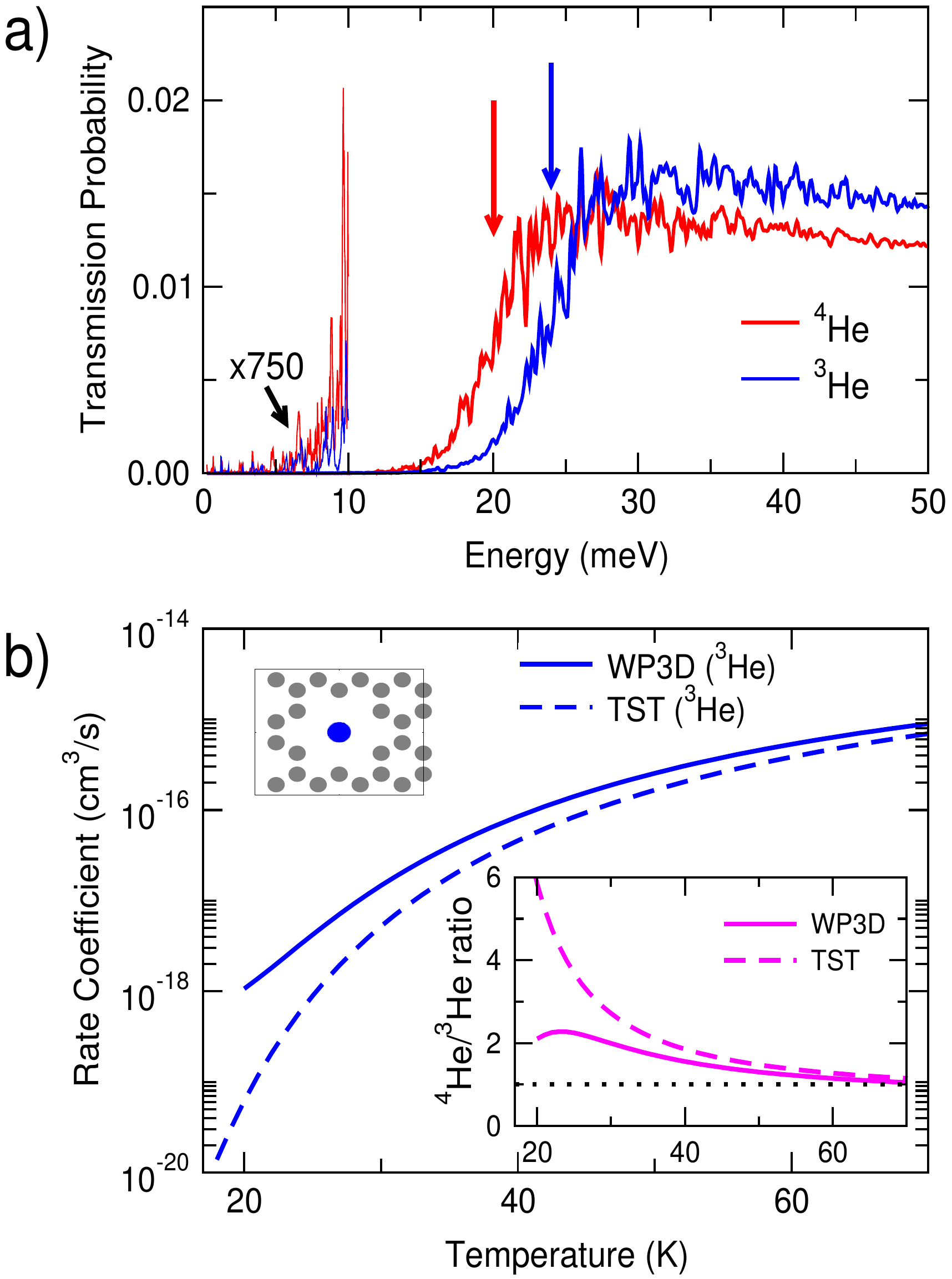}
\caption[] {a) transmission probabilities of $^4$He and $^3$He
  through the P7 membrane vs. kinetic energy of the atom. Red (blue) arrow 
  indicates the $^4$He ($^3$He) transmission threshold as   predicted by
  TST. For energies below 10 meV probabilities are shown magnified by a
  factor of 750. b) $^3He$ WP3D rate coefficients
  vs. temperature compared with TST estimations. P7 unit cell, including
  the TST effective area for $^3$He transmission,  
  are displayed in the upper-left inset. Finally, the $^4$He/$^3$He
  selectivity vs. temperature is presented in the lower-right inset.}
\label{fig4new}
\end{figure}

 He-P7 transmission probabilities are presented in Fig.\ref{fig4new}.a). As
expected, the probabilities rise at translational energies close to the TST
prediction given by the TS ground state level (20 and 24 meV for $^4$He
and $^3$He, respectively). However and in contrast to He-graphdiyne,
probabilities exhibit a multitude of peaks along the whole energy range. We
have checked that these structures are not due to any artifact in the
calculations. We believe that these peaks correspond to selective
adsorption resonances, a process that can be understood as a temporal trapping
of the incident wave into a bound state of the laterally averaged potential
while the motion along the parallel coordinates, ruled by the Bragg condition,
becomes faster for the sake of the conservation of
energy\cite{LJD:1936,SMiret:07,NuestroPRB:94}. Analysis of the wave packet
propagation supports this argument: it is noticed that after the main portion of 
the wave packet has been either reflected or transmitted by direct scattering,
a non-negligible fraction of this wave packet remains trapped along the
adsorption region ($\approx$ 3 \AA) for a long time while it is slowly 
decaying towards the reflection or the transmission regions. Resonance
structures are not seen in the He-graphdiyne transmission probabilities
probably because they are extremely narrow.  We plan to further study the role 
played by these resonances from simulations of the decaying of 
initially prepared adsorbed states\cite{NuestroPRB:94}.

WP3D rate coefficients for $^3$He-P7 as functions of temperature are reported
in Fig.\ref{fig4new}.b). Comparing with graphdiyne (Fig.\ref{fig2new}.b)), it
is worth noticing their large absolute values which are due to the lower
thresholds in the transmission probabilities. TST calculations have been also
performed for this system to test whether this theory can predict the
WP3D results. TST rate coefficients were obtained using Eq. \ref{ktst} where,
in this case, $f_{tunn}=1$, as the minimum energy path is barrier-less. The
$\gamma$ factor has been computed, as for graphdiyne, from the TS ground state
wave function and taking the same value for $f_{cut}$. The results are $\gamma
(TST)$= 0.0130 and 0.0144 for $^3$He and $^4$He, respectively, which are in
fairly good agreement with the heights of the plateaus reached for the
probabilities in the higher energy region (Fig.\ref{fig4new}.a)). 
However, TST rate coefficients do not agree quite well with the WP3D
calculations, especially at low temperatures where differences become of
almost two orders of magnitude. TST also underestimates the rate
coefficients of $^4$He-P7 (not shown) although to a lesser extent. As a
consequence, TST is unable to predict the qualitative behavior of the
$^4$He/$^3$He selectivity ratio, reported in the inset of
Fig.\ref{fig4new}.b). In fact, while TST predicts an increase of the 
$^4$He/$^3$He selectivity as temperature decreases (reaching a promising value
of six at 20 K), the accurate treatment gives a maximum of just 2.3 at about
23 K. Tunneling along non minimum energy paths (either direct or
mediated by the resonances\cite{KopinLiu:2012}) can be at the origin of the
larger values of the WP3D rate coefficients. The fact that discrepancies are
larger for the lighter isotope supports this argument. Therefore, it appears
that tunneling still operates for this system and competes with  ZPE
effects so that, after all, the quantum sieving is not as important as
initially expected based on simpler theories.

\vspace{0.5cm}

{\em Conclusion}. We have reported an accurate three-dimensional
wave packet approach for the study of the passage of atoms through nanoporous
one-atom-thick membranes. Results of simulations of the transmission of He
isotopes through graphdiyne and a leaky graphene model indicate the relevance
of quantum effects such as tunneling, transition state zero point energy and,
as a novelty, resonances. Transition state theory is 
found to be successful for He-graphdiyne but fails for the holey graphene
model.  This approach can be used as a reference in studies of the range of
validity of this and other approximate theories. Furthermore, it is
possible to extend this method to more complex systems such as diatoms or
bilayered membranes, for instance. We plan to work along some of these lines
in the near future.

\section*{Supporting Information}
Computational details of the three-dimensional wave packet calculations for
the transmission of atoms through periodic membranes.

\section*{Acknowledgments}
We thank Dr. Massimiliano Bartolomei for helpful discussions and a critical
reading of the manuscript. The work has been funded by Spanish MINECO grant FIS2013-48275-C2-1-P. 
 Allocation of computing time by CESGA (Spain) and support by 
the COST-CMTS Action CM1405 ``Molecules in Motion (MOLIM)'' 
are also  acknowledged.


\begin{mcitethebibliography}{48}
\providecommand*\natexlab[1]{#1}
\providecommand*\mciteSetBstSublistMode[1]{}
\providecommand*\mciteSetBstMaxWidthForm[2]{}
\providecommand*\mciteBstWouldAddEndPuncttrue
  {\def\EndOfBibitem{\unskip.}}
\providecommand*\mciteBstWouldAddEndPunctfalse
  {\let\EndOfBibitem\relax}
\providecommand*\mciteSetBstMidEndSepPunct[3]{}
\providecommand*\mciteSetBstSublistLabelBeginEnd[3]{}
\providecommand*\EndOfBibitem{}
\mciteSetBstSublistMode{f}
\mciteSetBstMaxWidthForm{subitem}{(\alph{mcitesubitemcount})}
\mciteSetBstSublistLabelBeginEnd
  {\mcitemaxwidthsubitemform\space}
  {\relax}
  {\relax}

\bibitem[Koenig et~al.(2012)Koenig, Wang, Pellegrino, and
  Bunch]{Nature-nano:2012}
Koenig,~S.~P.; Wang,~L.; Pellegrino,~J.; Bunch,~J.~S. Selective Molecular
  Sieving through Porous Graphene. \emph{Nature Nanotechnology} \textbf{2012},
  \emph{7}, 728--732\relax
\mciteBstWouldAddEndPuncttrue
\mciteSetBstMidEndSepPunct{\mcitedefaultmidpunct}
{\mcitedefaultendpunct}{\mcitedefaultseppunct}\relax
\EndOfBibitem
\bibitem[Jiao et~al.(2013)Jiao, Du, Hankel, and Smith]{Marlies:2013}
Jiao,~Y.; Du,~A.; Hankel,~M.; Smith,~S.~C. Modelling Carbon Membranes for Gas
  and Isotope Separation. \emph{Phys. Chem. Chem. Phys.} \textbf{2013},
  \emph{15}, 4832--4843\relax
\mciteBstWouldAddEndPuncttrue
\mciteSetBstMidEndSepPunct{\mcitedefaultmidpunct}
{\mcitedefaultendpunct}{\mcitedefaultseppunct}\relax
\EndOfBibitem
\bibitem[Huang et~al.(2015)Huang, Zhang, Li, and Shi]{review-JPCL:2015}
Huang,~L.; Zhang,~M.; Li,~C.; Shi,~G. Graphene-Based Membranes for Molecular
  Separation. \emph{J. Phys. Chem. Lett.} \textbf{2015}, \emph{6},
  2806--2815\relax
\mciteBstWouldAddEndPuncttrue
\mciteSetBstMidEndSepPunct{\mcitedefaultmidpunct}
{\mcitedefaultendpunct}{\mcitedefaultseppunct}\relax
\EndOfBibitem
\bibitem[Schrier(2010)]{Schrier:10}
Schrier,~J. Helium Separation Using Porous Graphene Membranes. \emph{J. Phys.
  Chem. Lett.} \textbf{2010}, \emph{1}, 2284--2287\relax
\mciteBstWouldAddEndPuncttrue
\mciteSetBstMidEndSepPunct{\mcitedefaultmidpunct}
{\mcitedefaultendpunct}{\mcitedefaultseppunct}\relax
\EndOfBibitem
\bibitem[Hauser and Schwerdtfeger(2012)Hauser, and Schwerdtfeger]{Hauser:2012}
Hauser,~A.~W.; Schwerdtfeger,~P. Nanoporous Graphene Membranes for Efficient
  $^3$He/$^4$He Separation. \emph{J. Phys. Chem. Lett.} \textbf{2012},
  \emph{3}, 209--213\relax
\mciteBstWouldAddEndPuncttrue
\mciteSetBstMidEndSepPunct{\mcitedefaultmidpunct}
{\mcitedefaultendpunct}{\mcitedefaultseppunct}\relax
\EndOfBibitem
\bibitem[Cho(2009)]{ScienceHe3:09}
Cho,~A. Helium-3 Shortage Could Put Freeze on Low-Temperature Research.
  \emph{Science} \textbf{2009}, \emph{326}, 778--779\relax
\mciteBstWouldAddEndPuncttrue
\mciteSetBstMidEndSepPunct{\mcitedefaultmidpunct}
{\mcitedefaultendpunct}{\mcitedefaultseppunct}\relax
\EndOfBibitem
\bibitem[Nuttall et~al.(2012)Nuttall, Clarke, and Glowacki]{Nature-He:2012}
Nuttall,~W.~J.; Clarke,~R.~H.; Glowacki,~B.~A. Resources: Stop Squandering
  Helium. \emph{Nature} \textbf{2012}, \emph{485}, 573--575\relax
\mciteBstWouldAddEndPuncttrue
\mciteSetBstMidEndSepPunct{\mcitedefaultmidpunct}
{\mcitedefaultendpunct}{\mcitedefaultseppunct}\relax
\EndOfBibitem
\bibitem[Cai et~al.(2012)Cai, Xing, and Zhao]{RSCAdv:2012}
Cai,~J.; Xing,~Y.; Zhao,~X. Quantum Sieving: Feasibility and Challenges for the
  Separation of Hydrogen Isotopes in Nanoporous Materiasls. \emph{RSC Advances}
  \textbf{2012}, \emph{2}, 8579--8586\relax
\mciteBstWouldAddEndPuncttrue
\mciteSetBstMidEndSepPunct{\mcitedefaultmidpunct}
{\mcitedefaultendpunct}{\mcitedefaultseppunct}\relax
\EndOfBibitem
\bibitem[Hauser et~al.(2012)Hauser, Schrier, and Schwerdtfeger]{Schrier:12}
Hauser,~A.~W.; Schrier,~J.; Schwerdtfeger,~P. Helium Tunneling through
  Nitrogen-Functionalized Graphene Pores: Pressure- and Temperature-Driven
  Approaches to Isotope Separation. \emph{J. Phys. Chem. C} \textbf{2012},
  \emph{116}, 10819--10827\relax
\mciteBstWouldAddEndPuncttrue
\mciteSetBstMidEndSepPunct{\mcitedefaultmidpunct}
{\mcitedefaultendpunct}{\mcitedefaultseppunct}\relax
\EndOfBibitem
\bibitem[Mandr{\`a} et~al.(2014)Mandr{\`a}, Schrier, and Ceotto]{ceotto:2014}
Mandr{\`a},~S.; Schrier,~J.; Ceotto,~M. Helium Isotope Enrichment by Resonant
  Tunneling through Nanoporous Graphene Bilayers. \emph{J. Phys. Chem. A}
  \textbf{2014}, \emph{118}, 6457--6465\relax
\mciteBstWouldAddEndPuncttrue
\mciteSetBstMidEndSepPunct{\mcitedefaultmidpunct}
{\mcitedefaultendpunct}{\mcitedefaultseppunct}\relax
\EndOfBibitem
\bibitem[Bartolomei et~al.(2014)Bartolomei, Carmona-Novillo, Hern{\'a}ndez,
  Campos-Mart{\'\i}nez, Pirani, and Giorgi]{ourJPCC2014}
Bartolomei,~M.; Carmona-Novillo,~E.; Hern{\'a}ndez,~M.~I.;
  Campos-Mart{\'\i}nez,~J.; Pirani,~F.; Giorgi,~G. Graphdiyne Pores: "Ad Hoc"
  Openings for Helium Separation Applications. \emph{J. Phys. Chem. C}
  \textbf{2014}, \emph{118}, 29966--29972\relax
\mciteBstWouldAddEndPuncttrue
\mciteSetBstMidEndSepPunct{\mcitedefaultmidpunct}
{\mcitedefaultendpunct}{\mcitedefaultseppunct}\relax
\EndOfBibitem
\bibitem[Lalitha et~al.(2015)Lalitha, Lakshmipathi, and Bhatia]{Lalitha2015}
Lalitha,~M.; Lakshmipathi,~S.; Bhatia,~S.~K. Defect-Mediated Reduction in
  Barrier for Helium Tunneling through Functionalized Graphene Nanopores.
  \emph{J. Phys. Chem. C} \textbf{2015}, \emph{119}, 20940--20948\relax
\mciteBstWouldAddEndPuncttrue
\mciteSetBstMidEndSepPunct{\mcitedefaultmidpunct}
{\mcitedefaultendpunct}{\mcitedefaultseppunct}\relax
\EndOfBibitem
\bibitem[Li et~al.(2015)Li, Qu, and Zhao]{Li:2015}
Li,~F.; Qu,~Y.; Zhao,~M. Efficient Helium Separation of Graphitic Carbon
  Nitride Membrane. \emph{Carbon} \textbf{2015}, \emph{95}, 51--57\relax
\mciteBstWouldAddEndPuncttrue
\mciteSetBstMidEndSepPunct{\mcitedefaultmidpunct}
{\mcitedefaultendpunct}{\mcitedefaultseppunct}\relax
\EndOfBibitem
\bibitem[Qu et~al.(2016)Qu, Li, Zhou, and Zhao]{Qu2016}
Qu,~Y.; Li,~F.; Zhou,~H.; Zhao,~M. Highly Efficient Quantum Sieving in Porous
  Graphene-like Carbon Nitride for Light Isotopes Separation. \emph{Scientific
  Reports} \textbf{2016}, \emph{6}, 19952\relax
\mciteBstWouldAddEndPuncttrue
\mciteSetBstMidEndSepPunct{\mcitedefaultmidpunct}
{\mcitedefaultendpunct}{\mcitedefaultseppunct}\relax
\EndOfBibitem
\bibitem[Beenakker et~al.(1995)Beenakker, Borman, and Krylov]{Beenakker:95}
Beenakker,~J. J.~M.; Borman,~V.~D.; Krylov,~S.~Y. Molecular Transport in
  Subnanometer Pores: Zero-Point Energy, Reduced Dimensionality and Quantum
  Sieving. \emph{Chem. Phys. Lett.} \textbf{1995}, \emph{232}, 379--382\relax
\mciteBstWouldAddEndPuncttrue
\mciteSetBstMidEndSepPunct{\mcitedefaultmidpunct}
{\mcitedefaultendpunct}{\mcitedefaultseppunct}\relax
\EndOfBibitem
\bibitem[Not()]{Note-zpe}
{We employ the term ``ZPE effects'' to refer in general to the effects due to
  the quantization of the states within the pore (identified as the transition
  state), and more especifically, to the in-pore ground state as it is the most
  populated state at low temperatures.}\relax
\mciteBstWouldAddEndPunctfalse
\mciteSetBstMidEndSepPunct{\mcitedefaultmidpunct}
{}{\mcitedefaultseppunct}\relax
\EndOfBibitem
\bibitem[Hankel et~al.(2011)Hankel, Zhang, Nguyen, Bhatia, Gray, and
  Smith]{hankel-pccp:2011}
Hankel,~M.; Zhang,~H.; Nguyen,~T.~X.; Bhatia,~S.~K.; Gray,~S.~K.; Smith,~S.~C.
  Kinetic Modelling of Molecular Hydrogen Transport in Microporous Carbon
  Materials. \emph{Phys. Chem. Chem. Phys.} \textbf{2011}, \emph{13},
  7834--7844\relax
\mciteBstWouldAddEndPuncttrue
\mciteSetBstMidEndSepPunct{\mcitedefaultmidpunct}
{\mcitedefaultendpunct}{\mcitedefaultseppunct}\relax
\EndOfBibitem
\bibitem[Kumar and Bathia(2005)Kumar, and Bathia]{Bathia:2005}
Kumar,~A. V.~A.; Bathia,~S.~K. Quantum Effect Induced Reverse Kinetic Molecular
  Sieving in Microporous Materials. \emph{Phys. Rev. Lett.} \textbf{2005},
  \emph{95}, 245901\relax
\mciteBstWouldAddEndPuncttrue
\mciteSetBstMidEndSepPunct{\mcitedefaultmidpunct}
{\mcitedefaultendpunct}{\mcitedefaultseppunct}\relax
\EndOfBibitem
\bibitem[Schrier and McClain(2012)Schrier, and McClain]{Schrier-cpl2012}
Schrier,~J.; McClain,~J. Thermally-Driven Isotope Separation across Nanoporous
  Graphene. \emph{Chem. Phys. Lett.} \textbf{2012}, \emph{521}, 118--124\relax
\mciteBstWouldAddEndPuncttrue
\mciteSetBstMidEndSepPunct{\mcitedefaultmidpunct}
{\mcitedefaultendpunct}{\mcitedefaultseppunct}\relax
\EndOfBibitem
\bibitem[Zhao et~al.(2006)Zhao, Villar-Rodil, Fletcher, and Thomas]{Zhao:06}
Zhao,~X.; Villar-Rodil,~S.; Fletcher,~A.~J.; Thomas,~K.~M. Kinetic Isotope
  Effect for H$_2$ and D$_2$ Quantum Molecular Sieving in Adsorption/Desorption
  on Porous Carbon Materials. \emph{J. Phys. Chem. B} \textbf{2006},
  \emph{110}, 9947--9955\relax
\mciteBstWouldAddEndPuncttrue
\mciteSetBstMidEndSepPunct{\mcitedefaultmidpunct}
{\mcitedefaultendpunct}{\mcitedefaultseppunct}\relax
\EndOfBibitem
\bibitem[Nguyen et~al.(2010)Nguyen, Jobic, and Bathia]{Nguyen2010}
Nguyen,~T.~X.; Jobic,~H.; Bathia,~S.~K. Microscopic Observation of Kinetic
  Molecular Sieving of Hydrogen Isotopes in a Nanoporous Material. \emph{Phys.
  Rev. Lett.} \textbf{2010}, \emph{105}, 085901\relax
\mciteBstWouldAddEndPuncttrue
\mciteSetBstMidEndSepPunct{\mcitedefaultmidpunct}
{\mcitedefaultendpunct}{\mcitedefaultseppunct}\relax
\EndOfBibitem
\bibitem[Hern{\'a}ndez et~al.(2015)Hern{\'a}ndez, Bartolomei, and
  Campos-Mert{\'\i}nez]{ourJPCA:2015}
Hern{\'a}ndez,~M.~I.; Bartolomei,~M.; Campos-Mert{\'\i}nez,~J. Transmission of
  Helium Isotopes through Graphdiyne Pores: Tunneling versus Zero Point Energy
  Effects. \emph{J. Phys. Chem. A} \textbf{2015}, \emph{119}, 10743--10749,
  (\small Note that, the rate coefficients computed and shown in Fig. 4 therein
  were erroneously multiplied by $(2\pi)^{-3/2}$. This error does not affect
  either the rest of the results nor the conclusions of the work.)\relax
\mciteBstWouldAddEndPuncttrue
\mciteSetBstMidEndSepPunct{\mcitedefaultmidpunct}
{\mcitedefaultendpunct}{\mcitedefaultseppunct}\relax
\EndOfBibitem
\bibitem[Li et~al.(2010)Li, Li, Liu, Guo, Li, and Zhu]{graphd-chemcomm:2010}
Li,~G.; Li,~Y.; Liu,~H.; Guo,~Y.; Li,~Y.; Zhu,~D. Architecture of Graphdiyne
  Nanoscale Films. \emph{Chem. Commun.} \textbf{2010}, \emph{46},
  3256--3258\relax
\mciteBstWouldAddEndPuncttrue
\mciteSetBstMidEndSepPunct{\mcitedefaultmidpunct}
{\mcitedefaultendpunct}{\mcitedefaultseppunct}\relax
\EndOfBibitem
\bibitem[Zhou et~al.(2015)Zhou, Gao, Liu, Xie, Yang, Zhang, Zhang, Liu, Li,
  Zhang, and Liu]{graphd-jacs:2015}
Zhou,~J.; Gao,~X.; Liu,~R.; Xie,~Z.; Yang,~J.; Zhang,~S.; Zhang,~G.; Liu,~H.;
  Li,~Y.; Zhang,~J.; Liu,~Z. Synthesis of Graphdiyne Nanowalls Using Acetylenic
  Coupling Reaction. \emph{J. Am. Chem. Soc.} \textbf{2015}, \emph{137},
  7596--7599\relax
\mciteBstWouldAddEndPuncttrue
\mciteSetBstMidEndSepPunct{\mcitedefaultmidpunct}
{\mcitedefaultendpunct}{\mcitedefaultseppunct}\relax
\EndOfBibitem
\bibitem[Jiao et~al.(2011)Jiao, Du, Hankel, Zhu, Rudolph, and
  Smith]{graphd-chemcomm:2011}
Jiao,~Y.; Du,~A.; Hankel,~M.; Zhu,~Z.; Rudolph,~V.; Smith,~S.~C. Graphdiyne: A
  Versatile Nanomaterial for Electronics and Hydrogen Purification. \emph{Chem.
  Commun.} \textbf{2011}, \emph{47}, 11843--11845\relax
\mciteBstWouldAddEndPuncttrue
\mciteSetBstMidEndSepPunct{\mcitedefaultmidpunct}
{\mcitedefaultendpunct}{\mcitedefaultseppunct}\relax
\EndOfBibitem
\bibitem[Cranford and Buehler(2012)Cranford, and Buehler]{nanoscalemitbis:2012}
Cranford,~S.~W.; Buehler,~M.~J. Selective Hydrogen Purification through
  Graphdiyne under Ambient Temperature and Pressure. \emph{Nanoscale}
  \textbf{2012}, \emph{4}, 4587--4593\relax
\mciteBstWouldAddEndPuncttrue
\mciteSetBstMidEndSepPunct{\mcitedefaultmidpunct}
{\mcitedefaultendpunct}{\mcitedefaultseppunct}\relax
\EndOfBibitem
\bibitem[Bartolomei et~al.(2014)Bartolomei, Carmona-Novillo, Hern{\'a}ndez,
  Campos-Mart{\'\i}nez, Pirani, Giorgi, and Yamashita]{jpclours:2014}
Bartolomei,~M.; Carmona-Novillo,~E.; Hern{\'a}ndez,~M.~I.;
  Campos-Mart{\'\i}nez,~J.; Pirani,~F.; Giorgi,~G.; Yamashita,~K. Penetration
  Barrier of Water through Graphynes' Pores: First-Principles Predictions and
  Force Field Optimization. \emph{J. Phys. Chem. Lett.} \textbf{2014},
  \emph{5}, 751--755\relax
\mciteBstWouldAddEndPuncttrue
\mciteSetBstMidEndSepPunct{\mcitedefaultmidpunct}
{\mcitedefaultendpunct}{\mcitedefaultseppunct}\relax
\EndOfBibitem
\bibitem[Balakrishnan et~al.(1997)Balakrishnan, Kalyanaraman, and
  Sathyamurthy]{reviewtdwp:97}
Balakrishnan,~N.; Kalyanaraman,~C.; Sathyamurthy,~N. Time-Dependent Quantum
  Mechanical Approach to Reactive Scattering and Related Processes. \emph{Phys.
  Rep.} \textbf{1997}, \emph{280}, 79--144\relax
\mciteBstWouldAddEndPuncttrue
\mciteSetBstMidEndSepPunct{\mcitedefaultmidpunct}
{\mcitedefaultendpunct}{\mcitedefaultseppunct}\relax
\EndOfBibitem
\bibitem[Balint-Kurti(2008)]{BKurti:08}
Balint-Kurti,~G.~G. Time-Dependent and Time-Independent Wavepacket Approaches
  to Reactive Scattering and Photodissociation Dynamics. \emph{Int. Rev. Phys.
  Chem.} \textbf{2008}, \emph{27}, 507--539\relax
\mciteBstWouldAddEndPuncttrue
\mciteSetBstMidEndSepPunct{\mcitedefaultmidpunct}
{\mcitedefaultendpunct}{\mcitedefaultseppunct}\relax
\EndOfBibitem
\bibitem[Yinnon and Kosloff(1983)Yinnon, and Kosloff]{Yinnon:83}
Yinnon,~A.~T.; Kosloff,~R. A Quantum-Mechanical Time-Dependent Simulation of
  the Scattering from a Stepped Surface. \emph{Chem. Phys. Lett.}
  \textbf{1983}, \emph{102}, 216--223\relax
\mciteBstWouldAddEndPuncttrue
\mciteSetBstMidEndSepPunct{\mcitedefaultmidpunct}
{\mcitedefaultendpunct}{\mcitedefaultseppunct}\relax
\EndOfBibitem
\bibitem[Sun et~al.(2014)Sun, Boutilier, Au, Poesio, Bai, Karnik, and
  Hadjiconstantinou]{CSun:2014}
Sun,~C.; Boutilier,~M. S.~H.; Au,~H.; Poesio,~P.; Bai,~B.; Karnik,~R.;
  Hadjiconstantinou,~N.~G. Mechanisms of Molecular Permeation through
  Nanoporous Graphene Membranes. \emph{Langmuir} \textbf{2014}, \emph{30},
  675--682\relax
\mciteBstWouldAddEndPuncttrue
\mciteSetBstMidEndSepPunct{\mcitedefaultmidpunct}
{\mcitedefaultendpunct}{\mcitedefaultseppunct}\relax
\EndOfBibitem
\bibitem[Lennard-Jones and Devonshire(1936)Lennard-Jones, and
  Devonshire]{LJD:1936}
Lennard-Jones,~J.~E.; Devonshire,~A.~F. Diffraction and Selective Adsorption of
  Atoms at Crystal Surfaces. \emph{Nature} \textbf{1936}, \emph{137},
  1069--1070\relax
\mciteBstWouldAddEndPuncttrue
\mciteSetBstMidEndSepPunct{\mcitedefaultmidpunct}
{\mcitedefaultendpunct}{\mcitedefaultseppunct}\relax
\EndOfBibitem
\bibitem[Sanz and Miret-Art{\'e}s(2007)Sanz, and Miret-Art{\'e}s]{SMiret:07}
Sanz,~A.~S.; Miret-Art{\'e}s,~S. Selective Adsorption Resonances: Quantum and
  Stochastic Approaches. \emph{Physics Reports} \textbf{2007}, \emph{451},
  37--154\relax
\mciteBstWouldAddEndPuncttrue
\mciteSetBstMidEndSepPunct{\mcitedefaultmidpunct}
{\mcitedefaultendpunct}{\mcitedefaultseppunct}\relax
\EndOfBibitem
\bibitem[Hern{\'a}ndez et~al.(1994)Hern{\'a}ndez, Campos-Mart{\'\i}nez,
  Miret-Art{\'e}s, and Coalson]{NuestroPRB:94}
Hern{\'a}ndez,~M.~I.; Campos-Mart{\'\i}nez,~J.; Miret-Art{\'e}s,~S.;
  Coalson,~R.~D. Lifetimes of Selective-Adsorption Resonances in Atom-Surface
  Elastic Scattering. \emph{Phys. Rev. B} \textbf{1994}, \emph{49},
  8300--8309\relax
\mciteBstWouldAddEndPuncttrue
\mciteSetBstMidEndSepPunct{\mcitedefaultmidpunct}
{\mcitedefaultendpunct}{\mcitedefaultseppunct}\relax
\EndOfBibitem
\bibitem[Feit et~al.(1982)Feit, Fleck, and Steiger]{SplitOperator}
Feit,~M.~D.; Fleck,~J.~A.; Steiger,~A. Solution of the Schr{\"o}dinger Equation
  by a Spectral Method. \emph{J. Comput. Phys.} \textbf{1982}, \emph{47},
  412--433\relax
\mciteBstWouldAddEndPuncttrue
\mciteSetBstMidEndSepPunct{\mcitedefaultmidpunct}
{\mcitedefaultendpunct}{\mcitedefaultseppunct}\relax
\EndOfBibitem
\bibitem[Kosloff and Kosloff(1983)Kosloff, and Kosloff]{K&K:1983}
Kosloff,~D.; Kosloff,~R. A Fourier Method Solution for the Time Dependent
  Schr{\"o}dinger Equation as a Tool in Molecular Dynamics. \emph{J. Comp.
  Phys.} \textbf{1983}, \emph{52}, 35--53\relax
\mciteBstWouldAddEndPuncttrue
\mciteSetBstMidEndSepPunct{\mcitedefaultmidpunct}
{\mcitedefaultendpunct}{\mcitedefaultseppunct}\relax
\EndOfBibitem
\bibitem[Heller(1975)]{Heller:75}
Heller,~E.~J. Time-Dependent Approach to Semiclassical Dynamics. \emph{J. Chem.
  Phys.} \textbf{1975}, \emph{62}, 1544--1555\relax
\mciteBstWouldAddEndPuncttrue
\mciteSetBstMidEndSepPunct{\mcitedefaultmidpunct}
{\mcitedefaultendpunct}{\mcitedefaultseppunct}\relax
\EndOfBibitem
\bibitem[Pernot and Lester(1991)Pernot, and Lester]{Pernot:91}
Pernot,~P.; Lester,~W.~A. Multidimensional Wave-Packet Analysis: Splitting
  Method for Time-Resolved Property Determination. \emph{Int. J. Quantum Chem.}
  \textbf{1991}, \emph{40}, 577--588\relax
\mciteBstWouldAddEndPuncttrue
\mciteSetBstMidEndSepPunct{\mcitedefaultmidpunct}
{\mcitedefaultendpunct}{\mcitedefaultseppunct}\relax
\EndOfBibitem
\bibitem[Miller(1974)]{Miller:74}
Miller,~W.~H. Quantum Mechanical Transition State Theory and a New
  Semiclassical Model for Reaction Rate Constants. \emph{J. Chem. Phys.}
  \textbf{1974}, \emph{61}, 1823--1834\relax
\mciteBstWouldAddEndPuncttrue
\mciteSetBstMidEndSepPunct{\mcitedefaultmidpunct}
{\mcitedefaultendpunct}{\mcitedefaultseppunct}\relax
\EndOfBibitem
\bibitem[Zhang and Zhang(1991)Zhang, and Zhang]{Zhang:91}
Zhang,~D.; Zhang,~J. Z.~H. Full-Dimensional Time-Dependent Treatment for
  Diatom-Diatom Reactions: The H$_2$+OH Reaction. \emph{J. Chem. Phys.}
  \textbf{1991}, \emph{101}, 1146--1156\relax
\mciteBstWouldAddEndPuncttrue
\mciteSetBstMidEndSepPunct{\mcitedefaultmidpunct}
{\mcitedefaultendpunct}{\mcitedefaultseppunct}\relax
\EndOfBibitem
\bibitem[di~Domenico et~al.(2001)di~Domenico, Hern{\'a}ndez, and
  Campos-Mart{\'\i}nez]{cplh2h2:01}
di~Domenico,~D.; Hern{\'a}ndez,~M.~I.; Campos-Mart{\'\i}nez,~J. A
  Time-Dependent Wave Packet Approach for Reaction and Dissociation in
  H$_2$+H$_2$. \emph{Chem. Phys. Lett.} \textbf{2001}, \emph{342},
  177--184\relax
\mciteBstWouldAddEndPuncttrue
\mciteSetBstMidEndSepPunct{\mcitedefaultmidpunct}
{\mcitedefaultendpunct}{\mcitedefaultseppunct}\relax
\EndOfBibitem
\bibitem[Truhlar and Kuppermann(1972)Truhlar, and Kuppermann]{Truhlar-jcp:72}
Truhlar,~D.~G.; Kuppermann,~A. Exact and Approximate Quantum Mechanical
  Reaction Probabilities and Rate Constants for the Collinear H + H$_2$
  Reaction. \emph{J. Chem. Phys.} \textbf{1972}, \emph{56}, 2232--2252\relax
\mciteBstWouldAddEndPuncttrue
\mciteSetBstMidEndSepPunct{\mcitedefaultmidpunct}
{\mcitedefaultendpunct}{\mcitedefaultseppunct}\relax
\EndOfBibitem
\bibitem[Garret and Truhlar(1979)Garret, and Truhlar]{Truhlar-jpc:79}
Garret,~B.~C.; Truhlar,~D.~G. Accuracy of Tunneling Corrections to Transition
  State Theory for Thermal Rate Constants of Atom Transfer Reactions. \emph{J.
  Phys. Chem.} \textbf{1979}, \emph{83}, 200--203\relax
\mciteBstWouldAddEndPuncttrue
\mciteSetBstMidEndSepPunct{\mcitedefaultmidpunct}
{\mcitedefaultendpunct}{\mcitedefaultseppunct}\relax
\EndOfBibitem
\bibitem[Wang et~al.(2016)Wang, Li, Yang, and Zhong]{Wang:2016}
Wang,~Y.; Li,~J.; Yang,~Q.; Zhong,~C. Two-Dimensional Covalent Triazine
  Framework Membrane for Helium Separation and Hydrogen Purification. \emph{ACS
  Appl. Mater. Interfaces} \textbf{2016}, \emph{8}, 8694--8701\relax
\mciteBstWouldAddEndPuncttrue
\mciteSetBstMidEndSepPunct{\mcitedefaultmidpunct}
{\mcitedefaultendpunct}{\mcitedefaultseppunct}\relax
\EndOfBibitem
\bibitem[Pirani et~al.(2008)Pirani, Brizi, Roncaratti, Casavecchia,
  Cappelletti, and Vecchiocattivi]{ILJ}
Pirani,~F.; Brizi,~S.; Roncaratti,~L.; Casavecchia,~P.; Cappelletti,~D.;
  Vecchiocattivi,~F. Beyond the Lennard-Jones Model: A Simple and Accurate
  Potential Function Probed by High Resolution Scattering Data Useful for
  Molecular Dynamics Simulations. \emph{Phys. Chem. Chem. Phys} \textbf{2008},
  \emph{10}, 5489--5503\relax
\mciteBstWouldAddEndPuncttrue
\mciteSetBstMidEndSepPunct{\mcitedefaultmidpunct}
{\mcitedefaultendpunct}{\mcitedefaultseppunct}\relax
\EndOfBibitem
\bibitem[Robeson(2008)]{Robeson:08}
Robeson,~L.~M. The Upper Bound Revisited. \emph{J. Membrane Sci.}
  \textbf{2008}, \emph{320}, 390--400\relax
\mciteBstWouldAddEndPuncttrue
\mciteSetBstMidEndSepPunct{\mcitedefaultmidpunct}
{\mcitedefaultendpunct}{\mcitedefaultseppunct}\relax
\EndOfBibitem
\bibitem[Liu(2012)]{KopinLiu:2012}
Liu,~K. Quantum Dynamical Resonances in Chemical Reactions: from A+BC to
  Polyatomic Systems. \emph{Adv. Chem. Phys.} \textbf{2012}, \emph{149},
  1--46\relax
\mciteBstWouldAddEndPuncttrue
\mciteSetBstMidEndSepPunct{\mcitedefaultmidpunct}
{\mcitedefaultendpunct}{\mcitedefaultseppunct}\relax
\EndOfBibitem
\end{mcitethebibliography}
\end{document}